\documentclass[3p,times,procedia]{elsarticle}
\usepackage{nupha_ecrc}

\volume{00}

\firstpage{1}

\journalname{Nuclear Physics A}

\runauth{}

\jid{nupha}

\jnltitlelogo{Nuclear Physics A}

\usepackage{amssymb}

\usepackage[figuresright]{rotating}

\newcommand{\jpsi}{J/\psi}
\newcommand{\ups}{\Upsilon}
\newcommand{\cQ}{{\cal Q}}
\newcommand{\sqrts}{\sqrt{s_{\rm NN}}}
\newcommand{\raa}{R_{\rm AA}}
\newcommand{\ie}{{\it i.e.}}
\newcommand{\eg}{{\it e.g.}}
\newcommand{\beq}{\begin{equation}}
\newcommand{\eeq}{\end{equation}}

\begin{document}

\begin{frontmatter}



\dochead{XXVIth International Conference on Ultrarelativistic Nucleus-Nucleus Collisions\\ (Quark Matter 2017)}

\title{Theoretical Perspective on Quarkonia from SPS via RHIC to LHC}


\author{R.~Rapp and X.~Du}
\address{$^1$Cyclotron Institute and Department of Physics and Astronomy,
Texas A\&M University, College Station, TX 77843-3366, USA}

\begin{abstract}
The objective of this paper is to assess the current theoretical understanding of the extensive set of quarkonium observables (for both charmonia and bottomonia) that have been attained in ultrarelativistic heavy-ion collisions over two orders
of magnitude in center-of-mass energy. We briefly lay out and compare the currently employed theoretical frameworks
and their underlying transport coefficients, and then analyze excitation functions of quarkonium yields to characterize
the nature of the varying production mechanisms. We argue that an overall coherent picture of suppression and regeneration mechanisms emerges which enables to deduce insights on the properties of the in-medium QCD force from SPS
via RHIC to LHC, and forms a basis for future quantitative studies.
\end{abstract}
­

\end{frontmatter}


\section{Introduction}
\label{sec_intro}
The fundamental force between a heavy quark ($Q$) and its anti-quark ($\bar Q$) in a
color singlet in the QCD vacuum is by now quantitatively established and can be represented
by a potential consisting of a short-distance Coulomb-type attraction and a long-range
linear ``confining" term,
\beq
V_{Q\bar Q} = - \frac{4}{3} \frac{\alpha_s}{r} + \sigma r \ .
\eeq
Here, $\alpha_s$  denotes the strong coupling constant and $\sigma$ the so-called string tension
arising from non-perturbative effects (\eg, gluonic condensates). The potential model has been
shown to emerge as a low-energy effective theory of QCD, it has been quantitatively confirmed
by lattice QCD (lQCD), and it yields a good description of spectroscopy for bound charmonia
($\Psi=\eta_c, \jpsi, \chi_c, \psi(2S)$, ...) and bottomonia ($Y =\eta_b, \ups(1S), \ups(2S)$,
...)~\cite{Brambilla:2010cs} .
The linear potential term turns out to be the main agent for the binding of all
quarkonia except for the ground-state $Y$ ($\eta_b$ and $\ups(1S)$); \eg, when switching off
the string term in the potential, eq. (1), the $\jpsi$ binding energy (commonly defined as
the energy gap to the $D\bar D$ threshold) drops by an order of magnitude.

Based on a well-calibrated QCD force in vacuum (and the spectrum it generates), we are provided with
an opportunity to deduce its modifications in hot and dense QCD matter through studying the in-medium
spectral properties of quarkonia. Quarkonium spectral functions in matter not only provide information on
the $Q\bar Q$ interactions, but also encode properties of open heavy flavor in medium, \eg, the
heavy-flavor (HF) diffusion coefficient or heavy-quark (HQ) susceptibilities, via suitable
low-energy and momentum limits.
In this way the spectral functions provide insights into generic properties of the quark-gluon 
plasma (QGP)  that ultimately result from the fundamental in-medium force.

\begin{figure}[t]
\begin{center}
\includegraphics[width=0.7\textwidth]{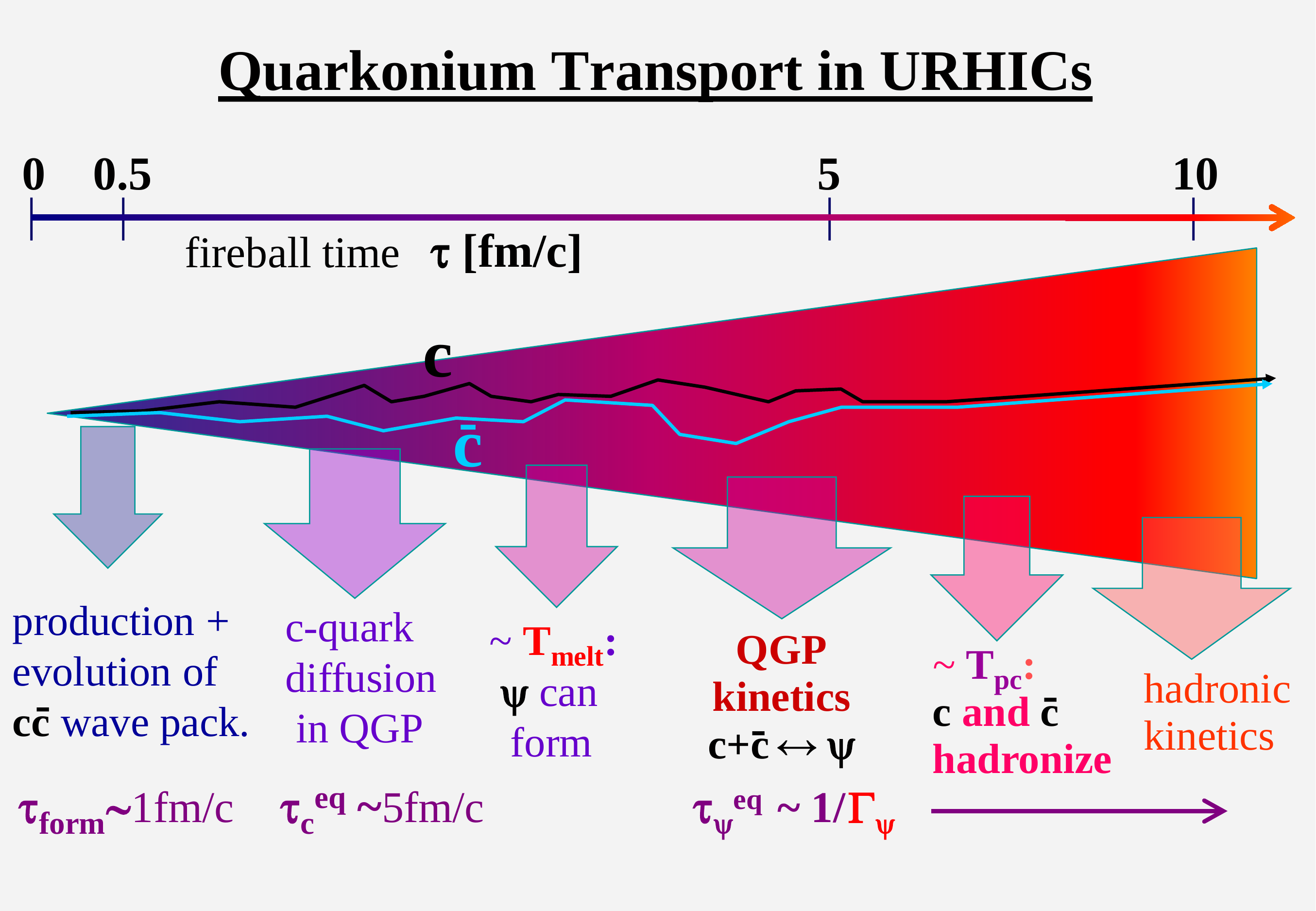}
\end{center}
\caption{Schematic time evolution of a correlated charm-anti-charm quark pair in an expanding fireball of
ultrarelativistic heavy-ion collisions, with timescales pertinent to charmonium transport (bottom line).
The original would-be $\jpsi$ first dissolves into $c$ and $\bar c$ which ultimately recombine again to
emerge as a charmonium in the final state.}
\label{fig_urhic}
\end{figure}

In the case of well-defined spectral peaks and mass thresholds, the quarkonium spectral functions directly
reflect the masses, binding energies and reaction rates (both elastic and inelastic) of the $Q\bar Q$ bound
states. However, if a spectral peak is about to melt, and/or if the scattering rates become very large (as is
expected for a strongly coupled medium), the spectral information does not lend itself to straightforward
interpretations. Model (or effective theory) calculations become necessary to interpret and apply the information
encoded in the spectral functions to experiment. This is a challenging task, but ample information
is available from lattice-QCD, \eg, Euclidean-time and spatial correlation functions, which provide strong
model constraints to control the time-like quantities needed for phenomenology. Ideally, one would then like
to infer the medium modifications of the QCD force from the experimental data on quarkonium production. For 
example, a successive melting of bound states according to their vacuum size could reveal how the force is 
progressively screened as temperature increases in a given collision system by changing collision centrality 
or energy. Such a ``force-meter" is quite different from the notion of using quarkonia as a thermometer.
One rather infers the temperature evolution of an ultrarelativistic heavy-ion collision (URHIC) from an
independent source (\eg, hydrodynamics, photon and/or dilepton spectra), and uses that as a reference
point to understand in-medium quarkonium properties. Regeneration processes complicate a straightforward
interpretation of quarkonium production yields, but they are an inevitable consequence of the re-emerging
bound states as the fireball cools. Detailed balance dictates that the same reactions that cause dissociation
are also operative for regeneration, although the latter is additionally affected by the individual heavy-quark
momentum distributions not being in thermal equilibrium. In turn, the in-medium interactions of heavy
quarks within quarkonia are key to understanding the latter's dissociation, \ie, open and hidden heavy flavor
in QCD matter (and URHICs) are intimately connected. While enlarging the scope of the problem, it will
ultimately strengthen the mutual consistency constraints between open and hidden HF kinetics. A sketch
of the different stages of the coupled quarkonium and heavy-quark/-hadron evolution through the fireball
expansion of URHICs is shown in Fig.~\ref{fig_urhic}, cf.~also the reviews
\cite{Rapp:2008tf,BraunMunzinger:2009ih,Kluberg:2009wc}.

In the following, we will briefly review basic theoretical ingredients to describe in-medium quarkonium
transport (Sec.~\ref{sec_tools}), analyze charmonium and bottomonium excitation functions for center-of-mass
collision energies $\sqrt{s_{\rm NN}}=0.017-5.02$\,TeV (Sec.~\ref{sec_excit}), and discuss two further
examples of in-medium QCD force strength probes (Sec.~\ref{sec_force}). We conclude in Sec.~\ref{sec_concl}.

\section{Theoretical Tools}
\label{sec_tools}
In URHICs, the production of HQ pairs, $Q\bar Q$, is expected to predominantly occur in primordial $NN$
collisions, being little affected in the subsequent fireball evolution with temperatures well below the
HQ mass, $T\ll m_Q$. In this situation, the thermal equilibrium number of a quarkonium state
($\cQ=\Psi, Y$)  of mass $m_Q$ , at temperature $T$ is given by
\beq
N_{\cQ}^{\rm eq}(T,\gamma_Q) = V_{\rm FB} \ d_{\cQ} \ \gamma_Q(T)^2
\int \frac{d^3p}{(2\pi)^3} \ f^B(m_{\cQ},T) \ ,
\label{Neq}
\eeq
where $d_{\cQ}$ denotes the spin degeneracy of $\cQ$ and $f^B$ the Bose distribution function. The HQ fugacity
factor, $\gamma_Q$ , is adjusted to match the equilibrium number of HF particles (hadrons or quarks) in the
fireball volume $V_{\rm FB}$ to the fixed number of HQ pairs, $N_{Q\bar Q}$.

In the statistical hadronization model~\cite{Gazdzicki:1999rk,Andronic:2006ky,BraunMunzinger:2009ih},
the production of charmonia and bottomonia is based on the thermal-equilibrium values of eq.~(\ref{Neq}).
They are evaluated at the chemical freezeout temperature $T_{\rm ch}\simeq160$\,MeV and baryon chemical
potential, $\mu_B^{\rm ch}$ (varying with collision energy), as determined from successful fits to
bulk-hadron production in URHICs over a wide range of collision energies. The underlying
idea is that a thermal QGP hadronizes and then rapidly falls out of chemical equilibrium as the inelastic
reaction rates drop with a large power of the particle densities.

In a more microscopic picture, transport approaches have been pursued to simulate the evolution of
quarkonia through an expanding fireball, along the lines sketched in Fig.~\ref{fig_urhic}.
Specifically, the semi-classical Boltzmann equation, schematically written as
\beq
p^\mu \partial_\mu f^{\cQ} = -E_p \Gamma_{\cQ} f^{\cQ} + E_p \beta \ ,
\eeq
describes the space-time evolution of the quarkonium distribution function, $f^{\cQ}$ , with a loss term
characterized by the rate $\Gamma_{\cQ}$ and a gain term with rate
$\beta$~\cite{Spieles:1999kp,Linnyk:2006ti,Liu:2010ej,Zhou:2014kka}.
Both rates are, in principle, based on the same micro-physics (transition matrix elements), but
$\beta$ also depends on the individual HQ distribution functions.
If the latter are in thermal equilibrium, the relation between gain and loss terms can be made more explicit by
integrating the Boltzmann equation over the spatial coordinates to obtain the rate equation (in the comoving
thermal frame),
\beq
\frac{dN_{\cQ}}{d\tau} = -\Gamma_{\cQ} \left[N_{\cQ} -N_{\cQ}^{\rm eq}\right] \ ,
\label{rate}
\eeq
which has also been deployed frequently to URHIC
phenomenology~\cite{Thews:2000rj,Grandchamp:2003uw,Chaudhuri:2008if,Zhao:2011cv,Song:2011xi,Strickland:2011aa,Ferreiro:2012rq}.
It shows that a single reaction rate, $\Gamma_{\cQ}$ , governs both suppression and regeneration processes,
driving the quarkonium number, $N_{\cQ}$, toward it's equilibrium value (the gain term is only active
if a quarkonium state can be supported at given temperature).
In this sense, $\Gamma_{\cQ}$ and $N_{\cQ}^{\rm eq}$ can be considered as transport parameters, where
the latter is the statistical-model value, eq.~(\ref{Neq}).
For the quarkonium reaction rate in the QGP, two main mechanisms have been considered:
gluo-dissociation~\cite{Bhanot:1979vb,Kharzeev:1994pz,Brambilla:2008cx,Liu:2013kkg},
$g+\cQ \to Q + \bar Q$ (also referred to a ``singlet-to-octet" mechanism) and inelastic parton
scattering~\cite{Grandchamp:2001pf,Laine:2006ns,Park:2007zza}, $p+\cQ \to p+ Q + \bar Q$ with
$p=q, \bar q, g$ (also referred to as ``quasi-free dissociation" or ``Landau damping" of the exchanged
gluon between $Q$ and $\bar Q$).
In weak coupling the latter, although naively of higher power in $\alpha_s$, takes over
from the former if the binding energy is much smaller than the Debye mass, $E_B \ll mD \sim gT$ .
In practice, with $g$$\simeq$2 and remnants of the confining force surviving up to $\sim2T_{\rm c}$ or so,
inelastic parton scattering turns out to take over already for $E_B \sim T$.

\begin{figure}[t]
\includegraphics[width=0.33\textwidth]{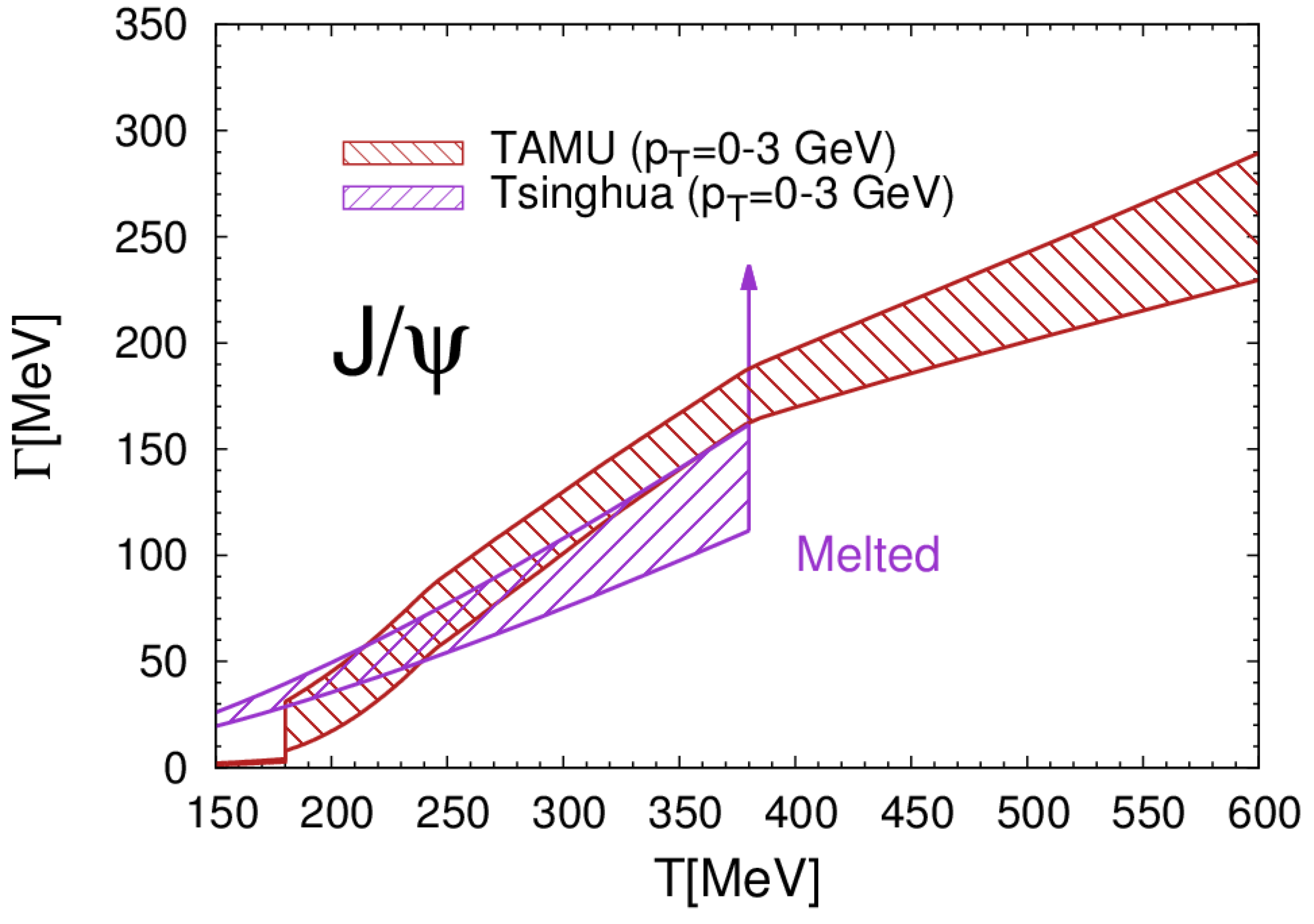}
\includegraphics[width=0.33\textwidth]{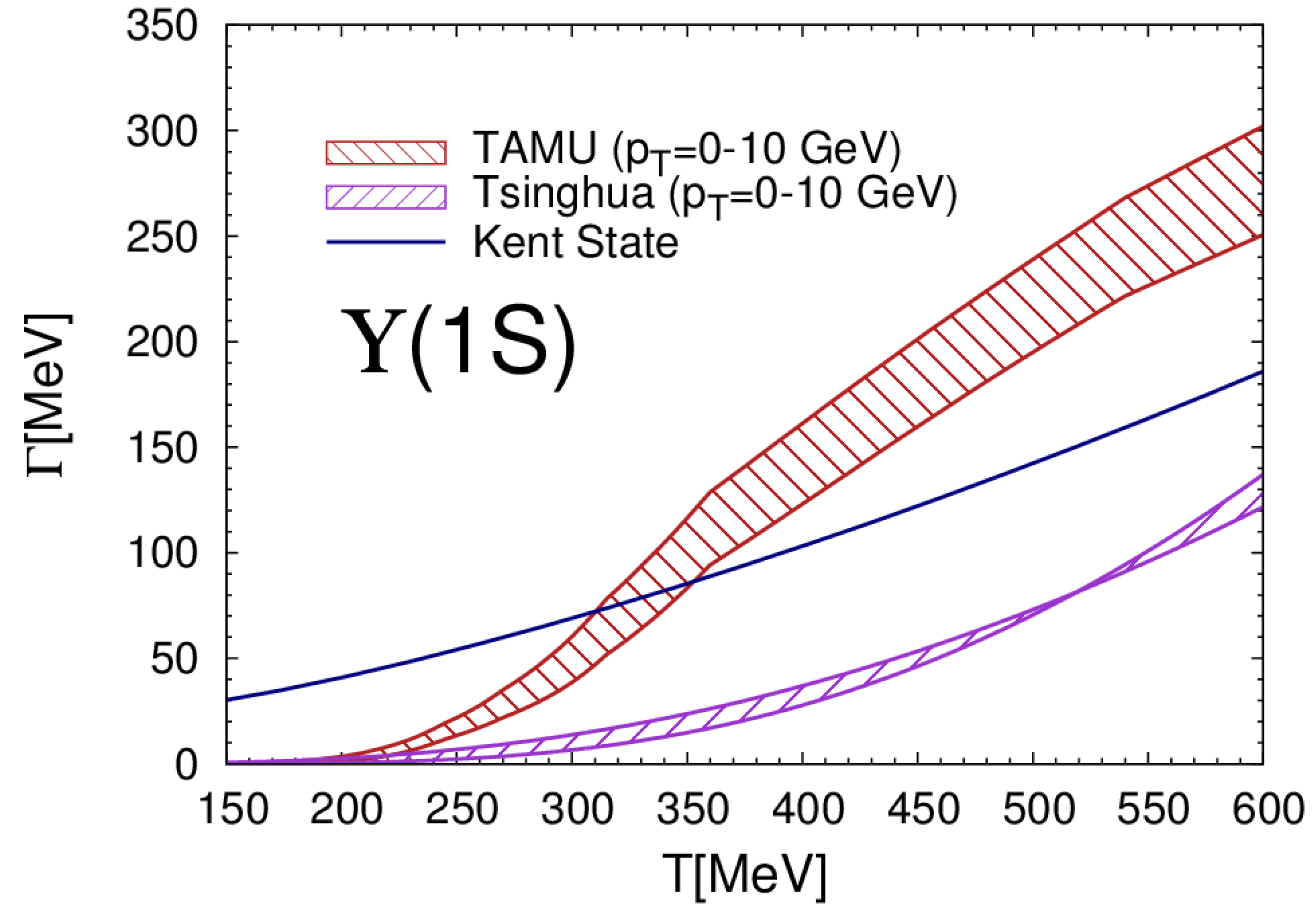}
\includegraphics[width=0.33\textwidth]{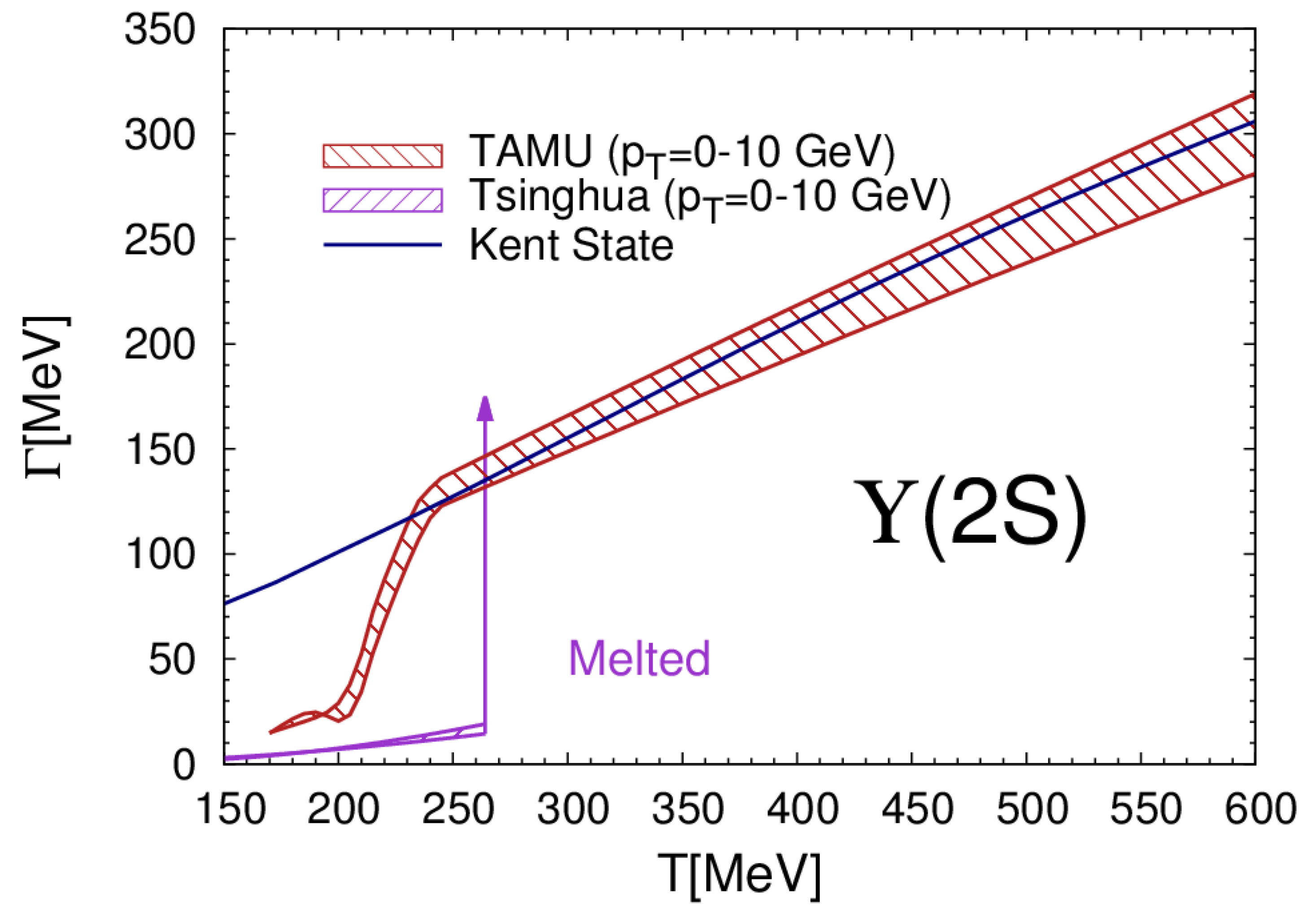}
\caption{Inelastic reaction rates of the $\jpsi$ (left), $\ups(1S)$ (middle) and $\ups(2S)$ (right)
in the QGP, comparing the results by the TAMU~\cite{Zhao:2011cv,Du:2017hss} (maroon bands),
Tsinghua~\cite{Zhou:2014kka,Liu:2010ej} (purple bands) and Kent-State~\cite{Strickland:2011aa} (blue 
lines) groups. The bands represent the indicated 3-momentum range (where applicable) most relevant for
the inclusive yields; an upward arrow indicates instantaneous suppression in the transport implementation,
while the other curves reach twice the individual heavy-quark width when the binding energy vanishes.}
\label{fig_gamma}
\end{figure}
In Fig.~\ref{fig_gamma} we compare inelastic quarkonium rates that figure in some of the transport
calculations used for phenomenology. For the $\jpsi$, one finds reasonable agreement between the
Tsinghua~\cite{Zhou:2014kka} and TAMU~\cite{Zhao:2011cv} groups (also with Ref.~\cite{Song:2011xi}),
although the underlying assumptions differ considerably (gluo-dissociation with vacuum binding vs.
quasi-free dissociation with in-medium binding). For the $\ups(1S)$ one
finds a larger spread between TAMU~\cite{Du:2017hss}, Tsinghua~\cite{Liu:2010ej} and
Kent-State~\cite{Strickland:2011aa} groups, both in magnitude and $T$ dependence.
For the $\ups(2S)$, which is strongly suppressed at the LHC~\cite{Chatrchyan:2012lxa}, the TAMU and Kent-State rates
agree rather well and gradually increase with $T$, while in the Tsinghua approach the suppression
is mostly realized through instantaneous melting for $T$$\gtrsim$260\,MeV.

\section{Quarkonium Excitation Functions}
\label{sec_excit}

The standard observable for quarkonia production in URHICs is the centrality dependence of their
nuclear modification factor, $\raa(N_{\rm part})$ (the yield normalized to the number expected
from an independent superposition of nucleon-nucleon ($NN$) collisions) for a given nucleus-nucleus
(AA) system at fixed energy, $\sqrt{s_{\rm NN}}$. A gradually increasing suppression with centrality
of up to a factor of $\sim$3 has been observed for $\jpsi$  production at SPS and RHIC energies and
an even larger [somewhat smaller] one for $\ups(2S)$ [$\ups(1S)$] at LHC energies, as a consequence
of the higher temperatures reached in more central collisions. On the contrary, the $\jpsi$
$\raa(N_{\rm part})$ at LHC energies quickly levels off at around 0.6-0.8 (depending on rapidity)
for $N_{\rm part}$$\gtrsim$100, strongly suggesting the prevalence of a new production mechanism that
was not readily identifiable at RHIC and SPS energies.

\begin{figure}[t]
\includegraphics[width=0.33\textwidth]{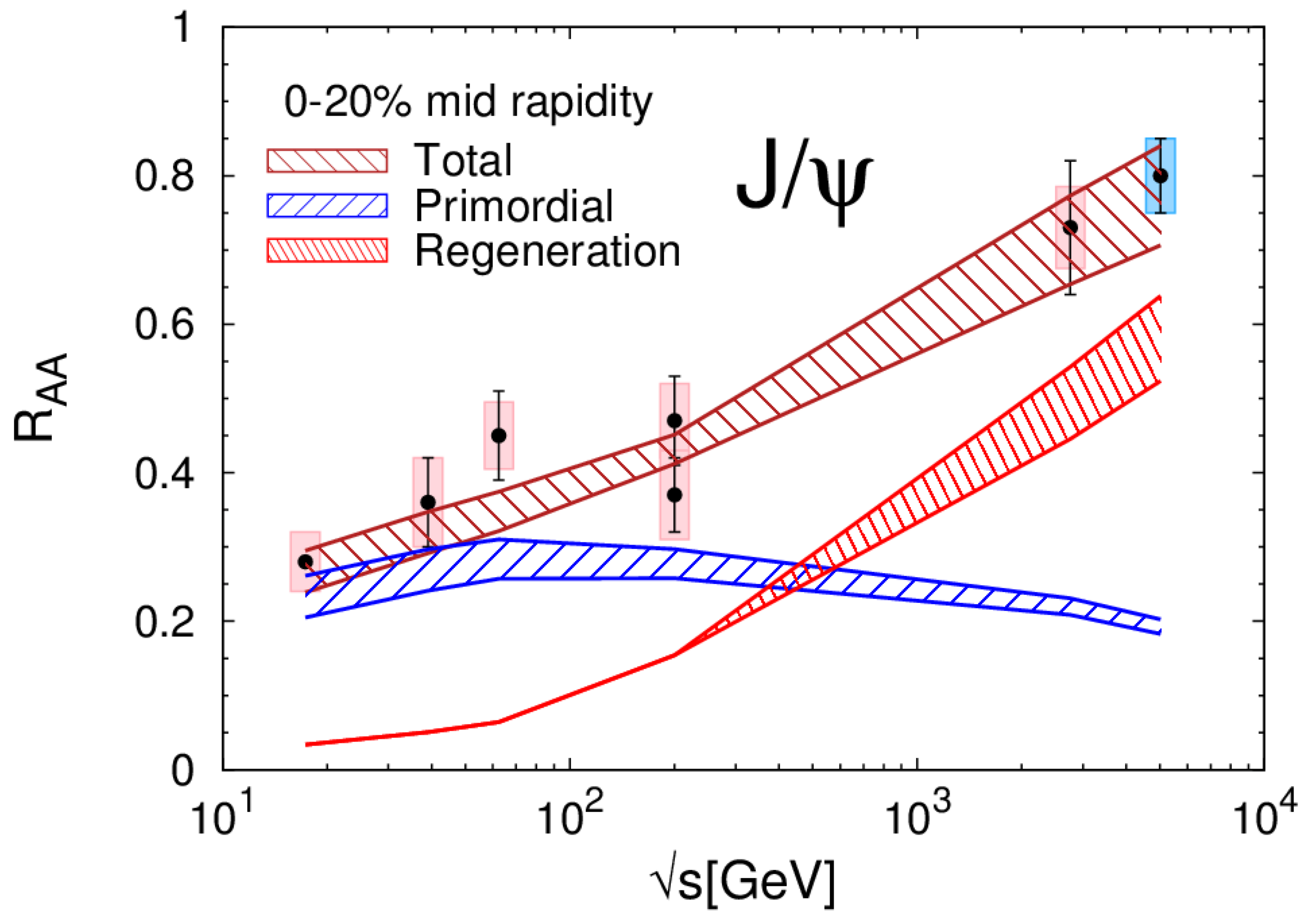}
\includegraphics[width=0.33\textwidth]{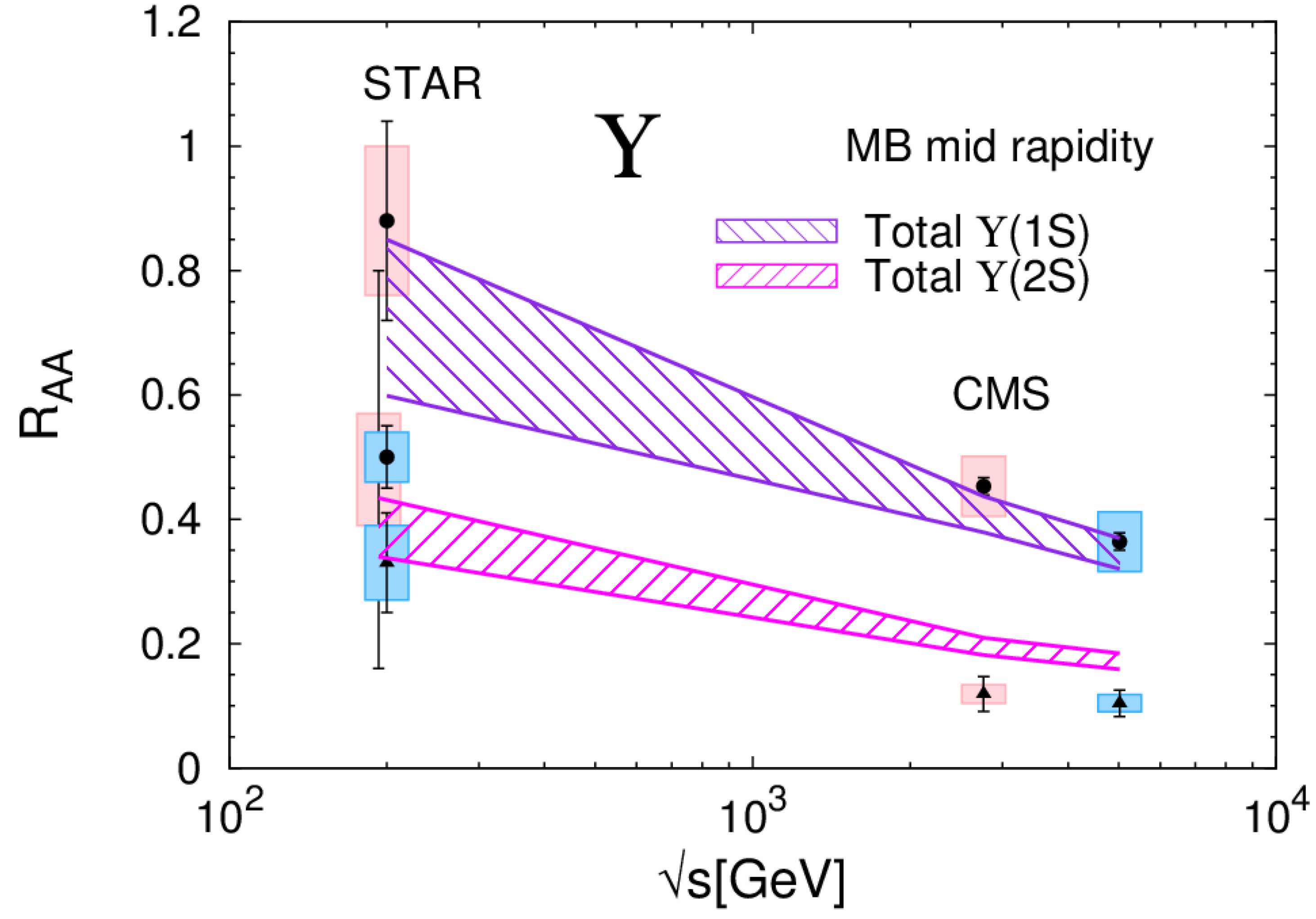}
\includegraphics[width=0.33\textwidth]{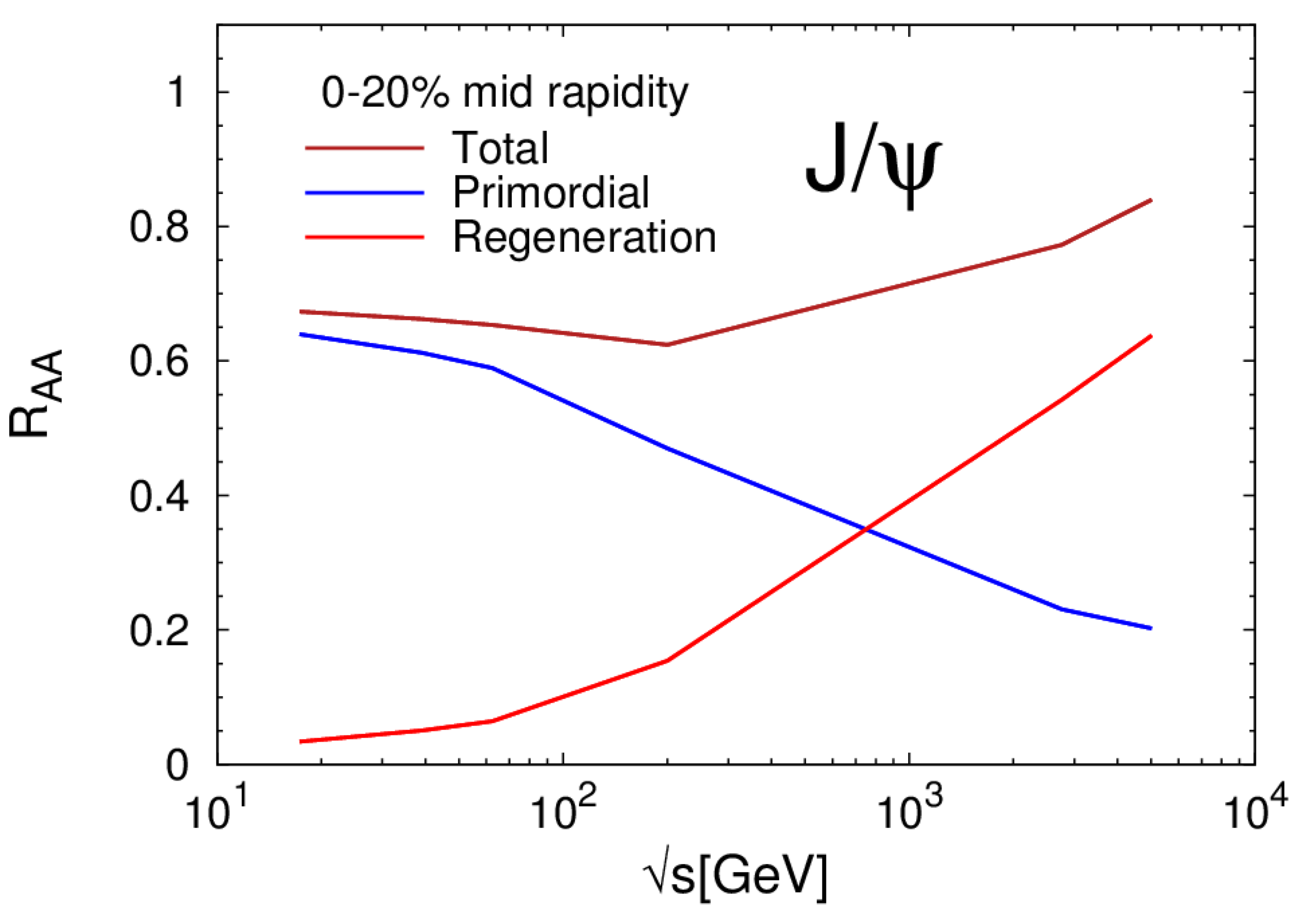}
\caption{Excitation functions for the inclusive $\jpsi$ (left panel), inclusive $\ups(1S,2S)$ (middle panel)
and $\jpsi$ corrected for CNM effects (right panel). Calculations for inclusive yields are compared to experimental
data~\cite{Alessandro:2004ap,Adare:2006ns,Adamczyk:2013tvk,Adamczyk:2016srz,Abelev:2013ila,Chatrchyan:2012lxa,Khachatryan:2016xxp},
including the newest preliminary ones released at this meeting (identified by blue systematic error
boxes)~\cite{Jimenez:qm2017,Ye:qm2017,Flores:qm2017}.
}
\label{fig_excit}
\end{figure}

As an alternative view of the production systematics, we compile in Fig.~\ref{fig_excit} the excitation function
of the $\raa(\sqrts)$ for inclusive $\jpsi$ and $\ups(1S,2S)$ in central and minimum-bias (MB) AA collisions,
respectively, at mid-rapidity from SPS (17\,GeV) via RHIC (39, 62, 200\,GeV) to the maximally available
energies at the LHC (2.76 and 5.02 TeV). The NA50~\cite{Alessandro:2004ap}, PHENIX~\cite{Adare:2006ns},
STAR~\cite{Adamczyk:2013tvk,Adamczyk:2016srz}, ALICE~\cite{Abelev:2013ila} and
CMS~\cite{Chatrchyan:2012lxa,Khachatryan:2016xxp} data are compared to theoretical calculations for both
$\Psi$ and $Y$ states in a common
theoretical framework, which solves a rate equation including suppression and regeneration with in-medium
binding energies (TAMU approach~\cite{Zhao:2011cv,Du:2017hss}; similar results are obtained in the Tsinghua
transport approach~\cite{Liu:2010ej,Zhou:2014kka}). The excitation function of the $\jpsi$ $\raa$ gradually
increases from about $\sim$0.3 at SPS to $\sim$0.8 at top LHC energy, interpreted as a strong increase in
regeneration, see left panel of Fig.~\ref{fig_excit}. On the contrary, both the
$\ups(1S)$ and $\ups(2S)$ $\raa$ decrease from RHIC to the LHC, see middle panel of Fig.~\ref{fig_excit}.
Despite their comparable vacuum binding energies, the $\ups(2S)$ $\raa$ at the LHC is $\sim$5 times smaller
than the one of the $\jpsi$! Regeneration is the only conceivable explanation for this. To better exhibit 
the effects of the hot medium on the $\jpsi$ $\raa$, we ``correct" the calculated values by taking out the
cold-nuclear-matter (CNM) effects, \ie, nuclear absorption of the nascent primordial $\jpsi$ and shadowing,
see right panel of Fig.~\ref{fig_excit}. It is reassuring to find that the ``primordial" component of the
$\jpsi$ $\raa$ excitation function now shows a behavior similar to the $\ups(2S)$ in the middle panel of
Fig.~\ref{fig_excit} (note that the $\jpsi$ still contains bottom feeddown, while the $\ups(2S)$ contains
regeneration, both at a near constant level of $\raa$$\simeq$1). Furthermore, the hot-matter $\raa$
of the $\jpsi$ reveals that its total suppression at the SPS is in large part due to CNM effects, caused
by a large nuclear-absorption cross section of $\sigma_{\rm abs}^{\jpsi}$$\simeq$7.5\,mb as extracted from
NA60 measurements in pA collisions at 17.3 GeV~\cite{Arnaldi:2010ky}. In fact, almost all of the inclusive
$\jpsi$'s hot-medium suppression is due to the (lack of) feeddown from (suppressed) excited states
($\chi_c$ and $\psi(2S)$), implying that the $\jpsi$ itself is actually rather robust within the QGP
formed at the SPS, where initial temperatures reach up to $T_0$$\simeq$240\,MeV as extracted, \eg,
from thermal dilepton radiation~\cite{Rapp:2014hha}. The suppression of the direct $\jpsi$ develops in the
RHIC energy
regime and reaches a factor of 5 or more at the LHC. The additional source of charmonia at the LHC is
further characterized by its concentration at low momenta, $p_T\lesssim m_{\jpsi}$, with a maximum of
the $\raa(p_T)$ close to zero~\cite{Adam:2015isa}, and a sizable elliptic flow~\cite{ALICE:2013xna},
as expected from theory~\cite{Zhou:2014kka,Zhao:2012gc}.
The softening of the $r_{\rm AA}\equiv \langle p_T^2\rangle_{\rm AA} / \langle p_T^2 \rangle_{pp}$,
introduced by the Tsinghua group~\cite{Zhou:2009vz}, from $\sim$1.5 at SPS~\cite{Abreu:2000xe} via
$\sim$1 at RHIC~\cite{Adare:2006ns} to $\sim$0.5 at LHC~\cite{Adam:2015rba} quantifies the transition
from primordial production with Cronin effect to regeneration
from a near-thermal source, respectively. These observations not only prove the presence of regeneration
processes, but imply vigorous reinteractions of charm and charmonia in the QGP, with large interaction rates
and $p_T$ spectra approaching thermalization, necessitating a strong coupling to the bulk medium.

Within the same theoretical framework, the observed suppression pattern in the bottomonium excitation functions 
(from RHIC to LHC), ordered by their binding energies, can be approximately explained. For the regeneration 
part, the question is not so much whether it exists but rather how significant it is. Current calculations 
suggest that it contributes at a level of $\sim$0.1 in the $\raa$ for both $\ups(1S)$ and $\ups(2S)$. In MB 
Pb-Pb collisions at the LHC, this amounts to a $\sim$25\% portion for the $\ups(1S)$ and more than 50\% for the 
$\ups(2S)$, which is appreciable. The $Y$ regeneration components are rather constant with centrality, and also
persist down to RHIC energies (although less significantly); the main reason for this small variation is that
bottom production is essentially in the canonical limit at both machines, \ie, no more that one $b \bar b$ pair
per unit of rapidity is produced in an AA collision. The TAMU calculations~\cite{Du:2017hss} shown in the middle
panel of Fig.~\ref{fig_excit} tend to overestimate the $\ups(2S)$ yields at the LHC, possibly due to an
overestimate of the regeneration part. Similar to the case of the $\jpsi$ , $p_T$ spectra can prove valuable
to disentangle primordial from regenerated bottomonia, although the less thermalized $b$-quark spectra entail
harder regenerated $Y$ spectra, which render a discrimination from the primordial spectra more challenging.
The newest $\ups(1S)$ $p_T$-spectra released at this meeting~\cite{Flores:qm2017} do indicate an intriguing
structure for $p_T \lesssim m_{\ups(1S)}$, in line with theory predictions
for a regeneration component~\cite{Du:2017hss}. An impressive number of new data points first released during
this meeting~\cite{Jimenez:qm2017,Ye:qm2017,Flores:qm2017} have also been included in Fig.~\ref{fig_excit};
they largely confirm the trends in the calculations.

\begin{figure}[t]
\begin{minipage}{5cm}
\vspace{0.5cm}
\hspace{-0.5cm}
\includegraphics[width=6.2cm]{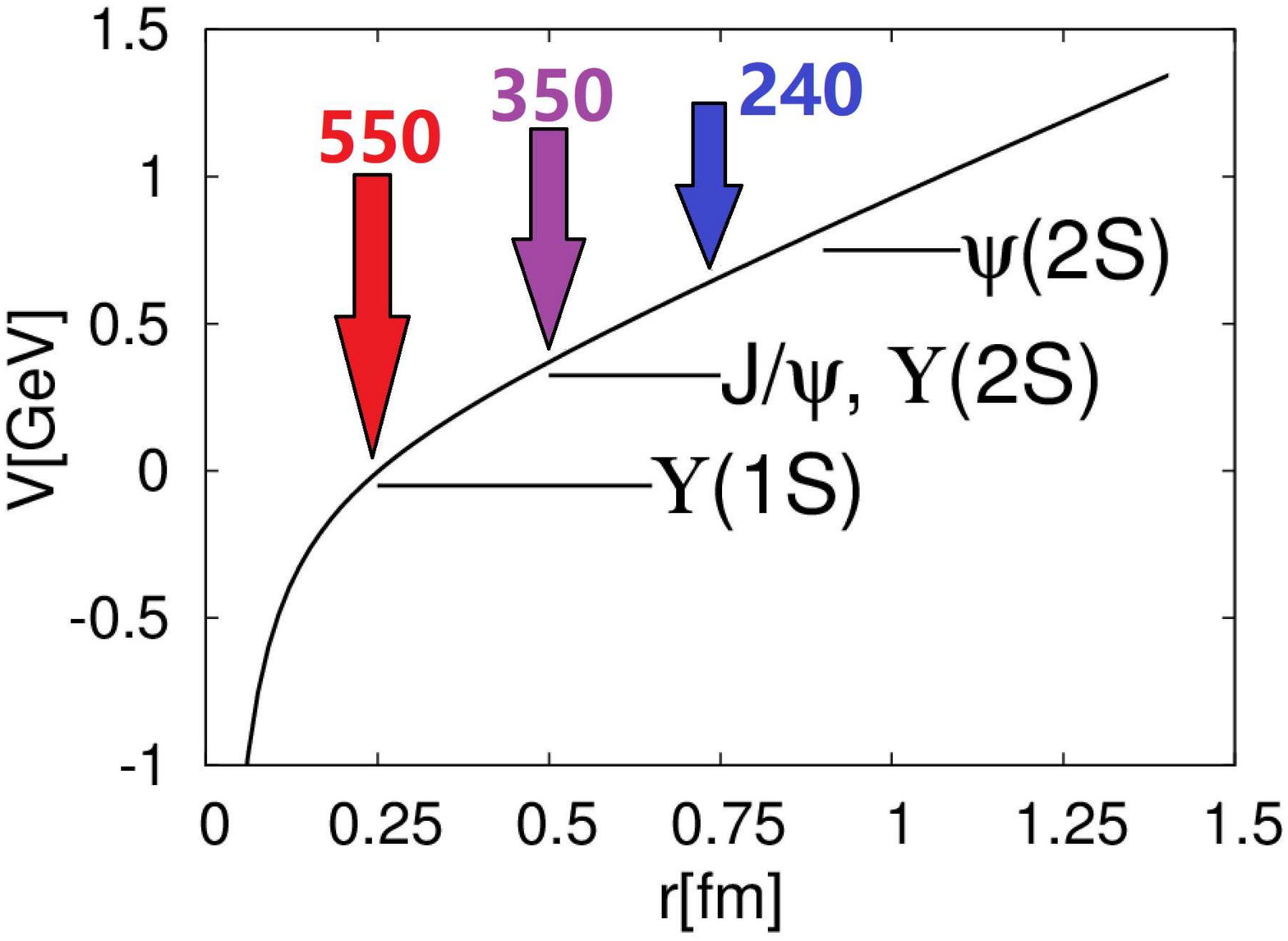}
\end{minipage}
\begin{minipage}{5cm}
\hspace{0.2cm}
\includegraphics[width=5.7cm]{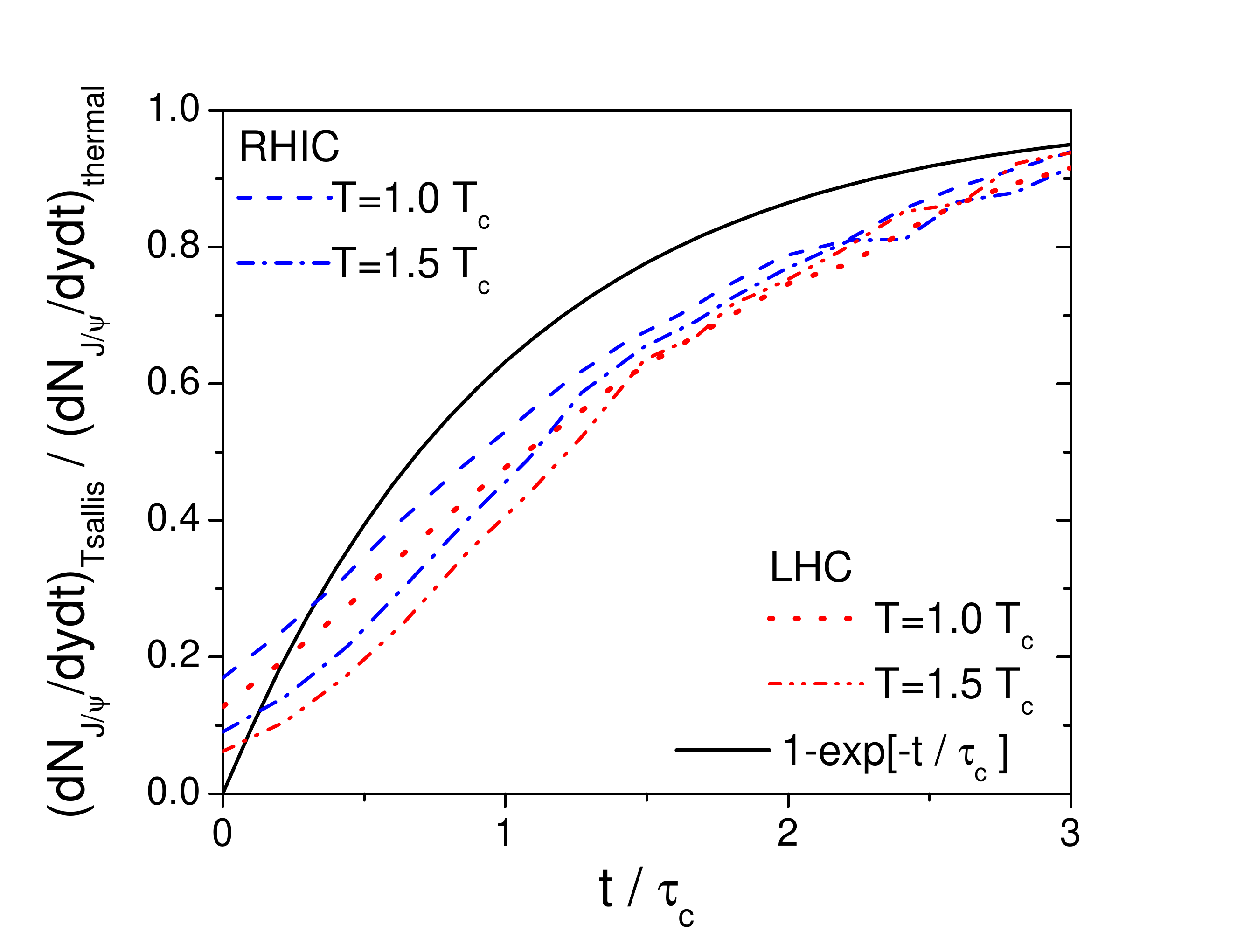}
\end{minipage}
\begin{minipage}{5cm}
\hspace{-0.3cm}
\includegraphics[width=6.1cm]{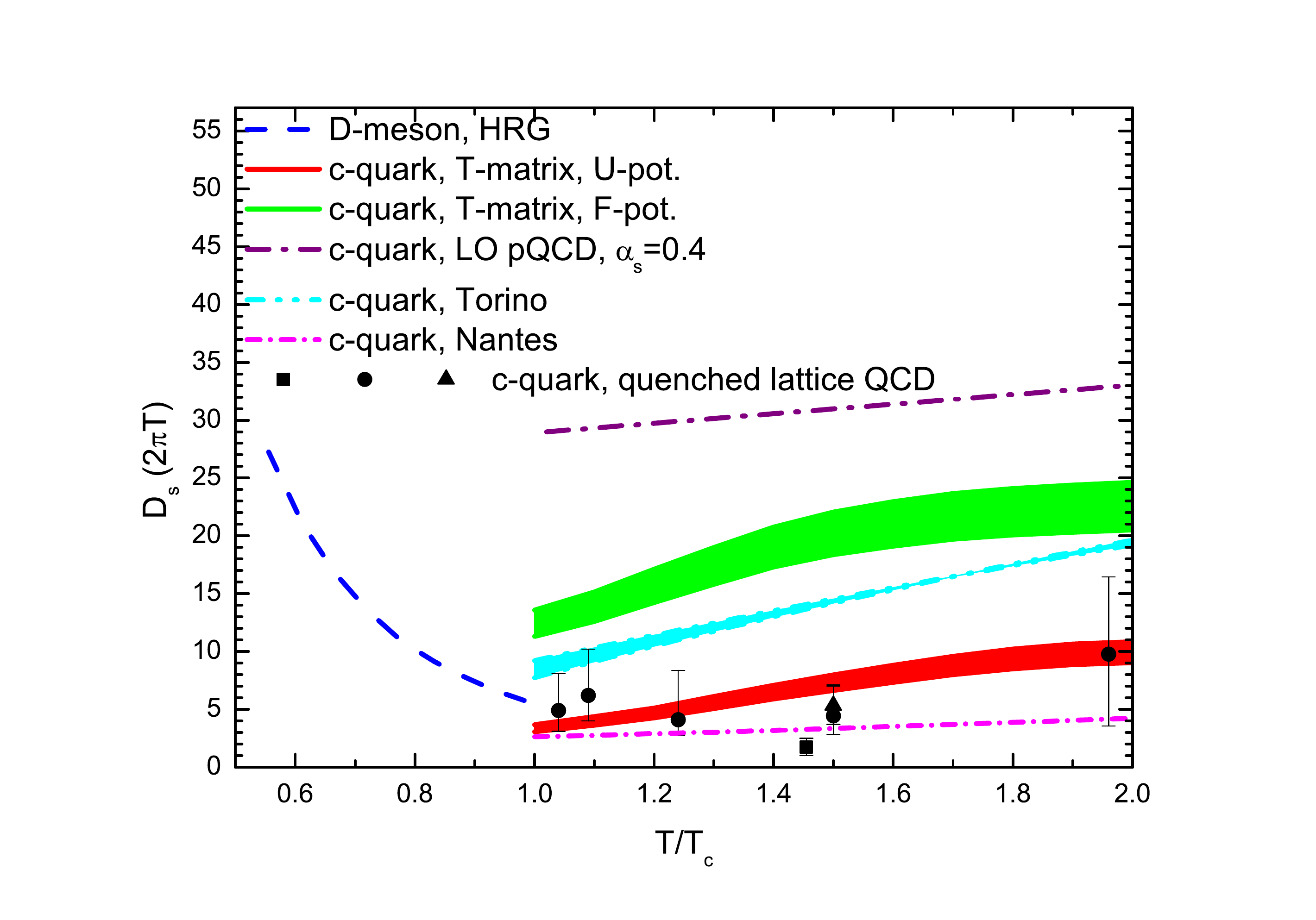}
\end{minipage}
\vspace{-0.3cm}
\caption{Heavy-flavor force strength probes of the QGP. Left panel: schematic conversion of the
quarkonium suppression pattern in URHICs to the screening of the QCD potential in medium at initial
temperatures realized in central Pb-Pb/Au-Au collisions at the SPS, RHIC and the LHC. Middle panel:
Dependence of the $\jpsi$ equilibrium number on reduced time in units of the HQ relaxation time
(figure adapted from Ref.~\cite{Song:2012at}). Right panel: compilation of calculations for the
spatial charm-quark diffusion coefficient (figure adapted from Ref.~\cite{Prino:2016cni}).
}
\label{fig_force}
\end{figure}

Let us now come back to the original objective of converting the quarkonium phenomenology in URHICs
into information on the in-medium QCD force. Based on the above discussion, we infer that
\begin{itemize}
\item Remnants of the confining force survive at the SPS [holding the $\jpsi$ together,
      but melting the $\psi(2S)$]
\item The confining force is screened at RHIC and the LHC [melting the $\jpsi$ and $\ups(2S)$]
\item The color-Coulomb force is screened at the LHC [strongly suppressing the $\ups(1S)$]
\item Thermalizing charm quarks recombine at the LHC [generating large $\jpsi$ yields].
\end{itemize}
These interpretations lead to the following hierarchy:
\beq
T_{\rm melt}[\psi(2S)] < T_0^{\rm SPS} < T_{\rm melt}[\jpsi,\ups(2S)]
\lesssim T_0^{\rm RHIC} < T_{\rm melt}[\ups(1S)] \lesssim T_0^{\rm LHC} \ .
\eeq
Extracting the initial temperatures from suitable bulk observables, \eg,
$T_0^{\rm SPS}$$\simeq$240\, MeV, $T_0^{\rm RHIC}$$\simeq$350\,MeV, $T_0^{\rm LHC}$$\simeq$550\,MeV,
and estimating pertinent screening radii as
$R_{\jpsi}^{\rm vac} < r_{\rm scr}[{\rm SPS}]$$\simeq$0.7\,fm $ < R_{\psi(2S)}^{\rm vac}$,
$R_{\ups(1S)}^{\rm vac} < r_{\rm scr}[{\rm RHIC}]$$\simeq$0.5\,fm $\lesssim R_{\jpsi}^{\rm vac}$,
and  $r_{\rm scr}[{\rm LHC}]$$\simeq$0.25\,fm$\lesssim R_{\rm \ups(1S)}^{\rm vac}$,
 we can relate them to medium effects on the
QCD force as illustrated in the left panel of Fig.~\ref{fig_force}. At the same time,
due to a large open-charm abundance at the LHC, the re-emerging confining force in the
later QGP phases inevitably regenerates charmonia.

In the above arguments, we also included the melting of the $\psi(2S)$ at the SPS, based on a factor of
up to $\sim$6 suppression in central Pb-Pb(17.3\,GeV) collisions~\cite{Alessandro:2006ju}. Interestingly,
the $\psi(2S)$ was also found to be appreciably suppressed in d-Au(0.2\,TeV)~\cite{Adare:2013ezl} and
p-Pb(5.02\,TeV)~\cite{Adam:2016ohd} collisions, well beyond expectations from CNM effects~\cite{Ferreiro:2014bia}.
The comover interaction model is able to explain this suppression (and the much smaller one for the $\jpsi$)
with effective interaction cross section, $\sigma_{\rm co}^{\Psi}$, extracted from reproducing SPS Pb-Pb
data. Converting these cross sections into dissociation widths,
$\Gamma_\Psi =\sigma_{\rm co}^\Psi n_{\rm co} v_{\rm rel}$, yields average values of 50-100\,MeV for the
$\psi(2S)$ and below 20\,MeV for the $\jpsi$ in dAu/pPb collisions. These are quite comparable to the thermal
widths discussed above, and, indeed, the suppression in small systems can also be understood in a
thermal-fireball framework~\cite{Du:2015wha}. The reaction rates from the comover and thermal
approaches thus support the formation of a ``medium" of duration 2-3\,fm in dAu/pPb collisions. The stronger
medium-induced suppression of the $\psi(2S)$ relative to the $\jpsi$ has important consequences for URHICs.
If the $\psi(2S)$ reaction rate is indeed active until lower temperatures than for the $\jpsi$, then $\psi(2S)$
regeneration should also operate at lower temperatures~\cite{Du:2015wha}. This could lead to interesting
effects in the $\psi(2S)$ $p_T$ spectra, due to a stronger collective flow imparted on the recombining charm
quarks in the later stages of the medium expansion.

\section{Force Strength Probes}
\label{sec_force}
The overall picture of quarkonium production in URHICs as outlined above generally supports a strong
coupling of $Q\bar Q$ bound states in medium, combining strong binding and vigorous chemistry (reaction rates).
Here we would like to discuss two additional, more specific aspects which relate to this picture.

The first is the impact of HQ thermalization on quarkonium regeneration. The primordially produced charm- 
and bottom-quark $p_T$-spectra from binary $NN$ collisions are significantly harder than thermal spectra
and thus provide unfavorable phase-space overlap for the formation of quarkonium bound states. The pertinent
reduction in the $\jpsi$ regeneration rate has been studied in Ref.~\cite{Song:2012at} by evolving initial
$c$-quark spectra at RHIC and the LHC toward their equilibrium value in a heat bath at fixed temperature,
cf.~middle panel in Fig.~\ref{fig_force}. The timescale of this evolution is given by the $c$-quark relaxation
time, $\tau_c$. The approach toward equilibrium is essentially universal, \ie, only depends on the ``reduced"
time, $t/\tau_c$, and not on temperature or initial conditions, and follows a relaxation time approximation,
${\cal R}(t)\simeq 1-\exp(-t/\tau_c)$. This factor has been introduced, \eg, into the TAMU transport approach,
via multiplication of the equilibrium limit in eq.~(\ref{rate})~\cite{Grandchamp:2003uw} (also included in
the calculations of Fig.~\ref{fig_excit}). To regenerate sufficient charmonia at the LHC, one needs a time
duration of at least $\tau_{\rm QGP} \simeq$~1-2~$\tau_c$ for charm quarks to reinteract with the medium,
implying $\tau_c\lesssim5$\,fm or so.
This directly relates to the force strength of the medium on slow-moving heavy quarks, typically quantified
by the HQ spatial diffusion coefficient, ${\cal D}_s=\tau_Q\frac{T}{m_Q}$. For charm quarks of mass
$m_c$=1.5\,GeV (which could be larger close to $T_{\rm c}$) and for a temperature range of $T$=0.2-0.3\,GeV,
the above constraint on the relaxation time translates into ${\cal D}_s (2\pi T) \lesssim$\,4-9, fully
compatible with current theoretical calculations with strong coupling and pertinent extractions from open
HF phenomenology in URHICs (see Fig.~\ref{fig_force} right and Ref.~\cite{Prino:2016cni} for a recent review).

The second example for a potentially direct force strength probe is the $\ups(1S)$. To bracket the medium
effects on its binding energy, several groups have calculated and compared results for the free ($F_{Q\bar Q}$)
vs. internal ($U_{Q\bar Q}$) HQ free energies as computed in lattice-QCD, as underlying potential. The two
quantities differ by an entropy term, $F_{Q\bar Q}(r; T) = U_{Q\bar Q} - T S_{Q\bar Q}$, which is operative
in the adiabatic (slow) limit (leading to $F$) but absent in the short-time limit (leading to $U$). The
former (latter) may thus be considered as a lower (upper) limit for the potential strength. In
Refs.~\cite{Strickland:2011aa,Emerick:2011xu}, the use of the $F_{Q\bar Q}$ was found to produce a suppression 
of the $\ups(1S)$ down to $\raa$$\simeq$0.1 in central Pb-Pb(2.76\,TeV), significantly below the CMS 
data~\cite{Chatrchyan:2012lxa}. On the other hand, with $U_{Q\bar Q}$ as potential much better agreement is 
found. At RHIC energies, this sensitivity is reduced as even the free energy provides significant binding for
temperatures $T$$\leq$300\,MeV. Interestingly, the $U_{Q\bar Q}$ potential is also much preferred in the
phenomenology of open HF in URHICs~\cite{Prino:2016cni} (recall Fig.~\ref{fig_force} right) and the related
question of $\jpsi$ regeneration discussed above.

\section{Conclusions}
\label{sec_concl}
The large amount of high-quality data emerging from systematic quarkonia measurements in URHICs
is creating a formidable challenge, but also a great opportunity, for unraveling the mechanisms for their
production. Theoretical descriptions using transport models for the space-time evolution of quarkonium
phase-space distributions turn out to provide a rather robust tool, with appreciable predictive power, to
capture the main features of the measured $\jpsi$, $\ups(1S)$ and $\ups(2S)$ production systematics, not
only as a function of centrality and transverse momentum, but also their excitation functions, now spanning a
factor of $\sim$300 in center-of-mass collision energies. We argued that this allows to disentangle suppression
and regeneration mechanisms for the $\jpsi$, yet to describe the gradually increasing suppression of the $Y$
states (where the role of regeneration remains to be scrutinized). We indicated how this information can be
used to determine quarkonium transport parameters and infer properties of the in-medium QCD force at the
different temperatures realized at the SPS, RHIC and the LHC. There is an encouraging degree of agreement
between transport models on the $\jpsi$ reaction rate, while the spread in the $Y$ rates requires further study.
We emphasized the intimate connection of in-medium quarkonia to the open HF sector, in particular the HF
diffusion coefficient. The latter directly reflects the coupling strength of individual low-momentum heavy
quarks to the medium, and as such bears on their ``quasi-free" reaction rates within a bound state, as well as
on the effectiveness of quarkonium regeneration (through their thermal relaxation).

Future efforts aimed at improving the theoretical precision of the transport framework need to tighten the
connections to the open HF sector (\eg, by implementing the explicit space-time dependence of HQ distributions
in the QGP), address the impact of quantum effects in the evolving quarkonium chemistry~\cite{Blaizot:2015hya,Bernard:2016spw,Brambilla:2016wgg}
and further develop the treatment of non-perturbative interactions near $T_{\rm c}$ that likely play a critical
role in understanding open HF observables. This might lead to larger quarkonium reaction rates than currently
employed in transport models, implying a faster approach toward chemical equilibrium of the quarkonium
yields, and thus coming closer to the equilibrium limit of the statistical model (as another transport parameter).
It has also become clear (not discussed here) that measurements of the open-charm (-bottom) cross
sections have to reach a 10\% precision level, to control predictions for regeneration yields at the 20(10)\%
level. Work in all these directions is well underway, providing promising perspectives for the future.
\\

\noindent
{\bf Acknowledgements}\\
We thank A. Andronic, B. Chen, E. Ferreiro, Y. Liu, T. Song, M. Strickland and P. Zhuang for their help
and discussions in preparing this presentation.
This work has been supported by U.S National Science Foundation under grant no. PHY-1614484.













\end{document}